\documentclass[twocolumn]{aastex6}
\usepackage{amssymb}
\usepackage{amsfonts}
\usepackage{amsmath}
\usepackage[varg]{txfonts}
\usepackage{natbib}
\usepackage{color}
\usepackage{graphicx}
\usepackage{hyperref}

\begin{document}

\let\oldint\int
\renewcommand{\int}[0]{\oldint\displaylimits}
\newcommand{\myvector}[1]{\mathbf{#1}}
\newcommand{\changed}[1]{{\color{red}#1}}

\title{Parallelized solution method of the three-dimensional gravitational
potential on the Yin-Yang grid}

\shorttitle{Parallelized solution method of the 3D gravitational potential on
the Yin-Yang grid}
\shortauthors{M.~Almanst{\"o}tter \emph{et al.}}

\author{Marius Almanst{\"o}tter\altaffilmark{1,2},
        Tobias Melson\altaffilmark{2},
        Hans-Thomas Janka\altaffilmark{2}, and
        Ewald M{\"u}ller\altaffilmark{2}}
\affil{\altaffilmark{1} Institut f\"ur Mathematik, Universit\"at Augsburg, 86135 Augsburg, Germany\\
\altaffilmark{2} Max-Planck-Institut f\"ur Astrophysik, Karl-Schwarzschild-Str.~1, 85748 Garching, Germany}

\begin{abstract}
We present a new method for solving the three-dimensional gravitational
potential of a density field on the Yin-Yang grid. Our algorithm is based on a
multipole decomposition and completely symmetric with respect to the two Yin-Yang
grid patches. It is particularly efficient on distributed-memory machines
with a large number of compute tasks, because the amount of data being
explicitly communicated is minimized.  All operations are performed on the
original grid without the need for interpolating data onto an auxiliary
spherical mesh.
\end{abstract}

\keywords{gravitation --- stars: general --- supernovae: general --- methods: numerical}

\section{Introduction}

Solving for Newtonian gravity in three dimensions (3D) is relevant in a large
number of astrophysical and geophysical problems. For instance, the propagation
of seismic waves on Earth \citep{Komatitsch2002}, the formation of planetesimals
in protoplanetary disks \citep{Simon2016}, shock propagation in protostellar
clouds \citep{Falle2017}, and the formation of the Earth's core
\citep{Mondal2018} are 3D situations, which require self-gravity to be taken
into account. One example, which is in the focus of our interest, are
core-collapse supernovae.

The explosion mechanism of core-collapse supernovae is one of the long-standing
riddles in stellar astrophysics. Thanks to growing supercomputing power,
3D neutrino-hydrodynamics simulations can nowadays be performed
to study the physical processes responsible for the onset of the explosion with
highly optimized codes on distributed-memory architectures
\citep{Takiwaki2012,Takiwaki2014,Melson2015b,Lentz2015,Melson2015a,Roberts2016,Mueller2017,Summa2018,Ott2018}.
The Garching group uses the \textsc{Prometheus-Vertex} package
\citep{Rampp2002}, which extends the finite-volume hydrodynamics module
\textsc{Prometheus} \citep{Fryxell1989} with a state-of-the-art neutrino
transport and interaction treatment. Its latest code version applies the
Yin-Yang grid \citep{Kageyama2004} -- a composite spherical mesh -- to
discretize the spatial domain.

Until now, self-gravity of the stellar plasma in 3D simulations has been treated
in spherical symmetry on an averaged density profile in \textsc{Prometheus},
because the gravitational potential is dominated by the spherical proto-neutron
star in the center. Since stellar collapse to neutron stars is only a mildly
relativistic problem, the Garching group applies Newtonian hydrodynamics but
uses a correction of the monopole of the gravitational potential
\citep{Marek2006}, which has been turned out to yield results that are well
compatible with fully relativistic calculations
\citep{Liebendoerfer2005a,Mueller2010}.
However, as 3D core-collapse supernova
simulations become more elaborate by taking more and more physical aspects into
account, it is highly desirable to include a realistic three-dimensional
gravitational potential to treat large-scale asymmetries correctly. This holds
true in particular in cases where the collapsing star develops a global
deformation, e.g.\ due to centrifugal effects in the case of rapid rotation.

\citet{Mueller2018} recently presented a method for solving Poisson's
equation on spherical polar grids using three-dimensional Fast Fourier
Transforms (FFT). Although their algorithm yields an accurate solution of the
gravitational potential even for highly aspherical density configurations, it is
currently not available for the Yin-Yang grid and its parallel efficiency needs
to be improved to serve our purposes.

In this work, we present a method for efficiently computing the gravitational
potential on the Yin-Yang grid based on the gravity solver of
\citet{Mueller1995}. Also \citet{Wongwathanarat2010} applied a three-dimensional
gravity solver on Yin-Yang data, however, their code was not parallelized for
distributed-memory systems. It relied on mapping the data to an auxiliary
spherical polar grid.  With our approach presented here, we compute the
gravitational potential on the Yin-Yang grid directly.

In Section \ref{sec:algorithm}, we will briefly summarize the algorithm
developed by \citet{Mueller1995} for 3D spherical grids. A method for its
efficient parallelization will be discussed in Section \ref{sec:parallel},
followed by a detailed explanation of the modifications for the Yin-Yang grid in
Section \ref{sec:yinyang}. Test calculations will be presented in Section
\ref{sec:tests}.


\section{Solving Poisson's equation on a spherical grid}
\label{sec:algorithm}

In this section, we briefly summarize the procedure for computing the
three-dimensional gravitational potential on a spherical polar grid as shown by
\citet{Mueller1995}.

Given the density distribution $\varrho(\myvector{r})$, the gravitational
potential at the location $\myvector{r}$ is determined by solving Poisson's equation,
which can be expressed in its integral form as
\begin{equation}
    \Phi(\myvector{r}) = -G\int_V \frac{\varrho(\myvector{r'})}{|\myvector{r}-\myvector{r'}|}
    \mathrm{d}^3\myvector{r'},
\end{equation}
where $G$ is the gravitational constant and $V$ denotes the whole computational
domain. In the following, we employ spherical coordinates and express spatial
vectors as $\myvector{r}=(r,\vartheta,\varphi)$.  A decomposition of
$|\myvector{r}-\myvector{r'}|^{-1}$ into spherical harmonics $Y_{\ell m}$ yields
\citep[][Eq.\ 8]{Mueller1995}
\begin{multline}
    \Phi(\myvector{r}) = -G \sum_{\ell=0}^{\infty} \sum_{m=-\ell}^\ell
    \frac{4\pi}{2\ell+1} Y_{\ell m}(\vartheta,\varphi) \,\times \\
    \times \left(\frac{1}{r^{\ell+1}} C_{\ell m}(r) +
    r^{\ell} D_{\ell m}(r)\right).
\end{multline}
In the latter equation, $C_{\ell m}$ and $D_{\ell m}$ are defined as
\begin{align}
    C_{\ell m}(r) &\coloneqq \int_0^{2\pi}\int_0^{\pi}
    \mathrm{d}\varphi'\mathrm{d}\vartheta' \sin\vartheta'\,
    \overline{Y_{\ell m}}(\vartheta',\varphi') \int_0^r\mathrm{d}r'
    (r')^{\ell+2}\varrho(\myvector{r'}), \\
    D_{\ell m}(r) &\coloneqq \int_0^{2\pi}\int_0^{\pi}
    \mathrm{d}\varphi'\mathrm{d}\vartheta' \sin\vartheta'\,
    \overline{Y_{\ell m}}(\vartheta',\varphi') \int_r^\infty\mathrm{d}r'
    (r')^{1-\ell}\varrho(\myvector{r'}),
\end{align}
where $\overline{Y_{\ell m}}$ are the complex conjugates of the spherical
harmonics.  After inserting their definition and rearranging the terms,
\citet{Mueller1995} wrote the gravitational potential as a sum of two
contributions: one for the gravitational potential inside a sphere of radius
$r$, $\Phi_{\mathrm{in}}^{(\ell m)}(\myvector{r})$, and a second for the potential
outside, $\Phi_{\mathrm{out}}^{(\ell m)}(\myvector{r})$,
\begin{align}
    \Phi(\myvector{r}) = -G\sum_{\ell=0}^{\infty}\sum_{m=0}^{\ell} N^{(\ell m)}
    P_\ell^m(\cos\vartheta) \left[\Phi_{\mathrm{in}}^{(\ell m)}(\myvector{r}) +
    \Phi_{\mathrm{out}}^{(\ell m)}(\myvector{r})\right],
    \label{eq:potential}
\end{align}
where $P_\ell^m$ are the associated Legendre polynomials.  The integrals for
these inner and outer contributions can be expressed as
\begin{multline}
    \Phi_{\textrm{in}}^{(\ell m)}(\myvector{r}) = \frac{1}{r^{\ell+1}}
    \int_0^{2\pi}\int_0^{\pi}\mathrm{d}\varphi'\mathrm{d}\vartheta'\sin\vartheta'
    P_\ell^m(\cos\vartheta')\,\times \\
    \times \cos\!\left(m(\varphi-\varphi')\right) \int_0^r\mathrm{d}r'
    {(r')}^{\ell+2}\varrho(\myvector{r'})
    \label{eq:phiin_int}
\end{multline}
and
\begin{multline}
    \Phi_{\textrm{out}}^{(\ell m)}(\myvector{r}) = {r}^\ell
    \int_0^{2\pi}\int_0^{\pi}\mathrm{d}\varphi'\mathrm{d}\vartheta'\sin\vartheta'
    P_\ell^m(\cos\vartheta')\, \times \\
    \times \cos\!\left(m(\varphi-\varphi')\right) \int_r^\infty\mathrm{d}r'
    {(r')}^{1-\ell}\varrho(\myvector{r'}).
    \label{eq:phiout_int}
\end{multline}
The normalization factor in Eq.\ (\ref{eq:potential}) is given by
\begin{equation}
    \mathcal{N}^{(\ell m)} \coloneqq \frac{(\ell-m)!}{(\ell+m)!}\frac{2}{\delta^{(m)}},
\end{equation}
with
\begin{equation}
    \delta^{(m)} \coloneqq
    \begin{cases} 2, & \mathrm{if}~m = 0, \\
        1, & \mathrm{if}~m > 0.
    \end{cases}
\end{equation}

To discretize these equations, we divide the spatial domain into $n_R \times
n_\vartheta \times n_\varphi$ grid cells, each spanning from $(r_i^-,
\vartheta_j^-, \varphi_k^-)$ to $(r_i^+, \vartheta_j^+, \varphi_k^+)$ with
$i=1,\ldots,n_R$, $j=1,\ldots,n_\vartheta$, and $k=1,\ldots,n_\varphi$. The
finite-volume method as being used in the \textsc{Prometheus} code assumes that
in each cell, the density is given as a cell average, i.e.,
\begin{align}
    \varrho(\myvector{r}) = \varrho_{ijk},
\end{align}
for $r_i^- \leq r \leq r_i^+$, $\vartheta_j^- \leq \vartheta \leq \vartheta_j^+$,
and $\varphi_k^- \leq \varphi \leq \varphi_k^+$.  This assumption allows for
simplifying Eqs.\ (\ref{eq:phiin_int}) and (\ref{eq:phiout_int}), which can then be
written as
\begin{align}
    \Phi_{\textrm{in}}^{(\ell m)}(\myvector{r}) &= \frac{1}{r^{\ell+1}}
    \sum_{i=1}^{n_r}\sum_{j=1}^{n_{\vartheta}}\sum_{k=1}^{n_{\varphi}}
    \varrho_{ijk}\, \mathcal{R}_{\textrm{in},i}^{(\ell)}\,
    \mathcal{T}_j^{(\ell m)}\, \mathcal{F}_k^{(m)}(\varphi),
    \label{eq:phiin_tot} \\
    \Phi_{\textrm{out}}^{(\ell m)}(\myvector{r}) &= r^\ell
    \sum_{i=n_r+1}^{n_R}\sum_{j=1}^{n_{\vartheta}}\sum_{k=1}^{n_{\varphi}}
    \varrho_{ijk}\, \mathcal{R}_{\textrm{out},i}^{(\ell)}\,
    \mathcal{T}_j^{(\ell m)}\, \mathcal{F}_k^{(m)}(\varphi),
    \label{eq:phiout_tot}
\end{align}
with
\begin{align}
    \mathcal{F}_k^{(m)}(\varphi) \coloneqq \cos(m\varphi)\,\mathcal{C}_k^{(m)} +
    \sin(m\varphi)\,\mathcal{S}_k^{(m)}.
\end{align}
In our chosen coordinates, the gravitational potential is computed at the cell
interfaces. The radial index $n_r$ introduced above is equal to the cell index
$i$ if $r = r_i^+$. The two integrals $\mathcal{C}_k^{(m)}$ and
$\mathcal{S}_k^{(m)}$ can be evaluated analytically,
\begin{align}
    \mathcal{C}_k^{(m)} &\coloneqq \int_{\varphi_k^{-}}^{\varphi_k^{+}}\mathrm{d}\varphi'
    \cos\!\left(m\varphi'\right)\nonumber \\
    &= \begin{cases}
        \varphi_k^{+} - \varphi_k^{-} = \Delta\varphi_k, & \mathrm{if}~m=0, \\
        \frac{1}{m}\left[\sin\!\left(m\varphi_k^{+}\right)-
        \sin\!\left(m\varphi_k^{-}\right)\right], & \mathrm{if}~m>0,
       \end{cases}
\end{align}
and
\begin{align}
    \mathcal{S}_k^{(m)} &\coloneqq \int_{\varphi_k^{-}}^{\varphi_k^{+}}\mathrm{d}\varphi'
    \sin\!\left(m\varphi'\right)\nonumber \\
    &= \begin{cases}
        0, & \mathrm{if}~m=0, \\
        \frac{1}{m}\left[\cos\!\left(m\varphi_k^{-}\right)-
        \cos\!\left(m\varphi_k^{+}\right)\right], & \mathrm{if}~m>0.
       \end{cases}
\end{align}
The radial integrals in Eqs.\ (\ref{eq:phiin_tot}) and (\ref{eq:phiout_tot}) can also be computed directly,
\begin{align}
    \mathcal{R}_{\textrm{in},i}^{(\ell)} &\coloneqq \int_{r_i^{-}}^{r_i^{+}}\mathrm{d}r'{(r')}^{\ell+2}
    = \frac{1}{\ell+3}\left((r_i^{+})^{\ell+3}-(r_i^{-})^{\ell+3}\right),
    \label{rin} \\
    \mathcal{R}_{\textrm{out},i}^{(\ell)} &\coloneqq \int_{r_i^{-}}^{r_i^{+}}\mathrm{d}r'{(r')}^{1-\ell}
    = \begin{cases}
        \ln(r_i^{+}) - \ln(r_i^{-}), & \mathrm{if}~\ell = 2, \\
        \frac{1}{2-\ell}\left[(r_i^{+})^{2-\ell}-(r_i^{-})^{2-\ell}\right], &
        \mathrm{if}~\ell\neq 2.
      \end{cases}
\end{align}
The remaining integrals in Eqs.\ (\ref{eq:phiin_tot}) and (\ref{eq:phiout_tot}),
\begin{equation}
    \mathcal{T}_j^{(\ell m)} \coloneqq \int_{\vartheta^{-}_j}^{\vartheta^{+}_j}\mathrm{d}\vartheta'
    \sin\vartheta' P_\ell^m(\cos\vartheta'),
    \label{eq:tlm}
\end{equation}
can be evaluated efficiently and analytically, i.e.\ without numerical
integration errors, using recurrence relations (see Appendix \ref{appendix}).

Solving Eq.\ (\ref{eq:potential}) numerically requires setting an upper bound
for the summation over $\ell$, which we denote as $\ell_\mathrm{max}$. We will
discuss the choice of $\ell_\mathrm{max}$ below in Section \ref{sec:discussion}.

\section{Parallelization of the method}
\label{sec:parallel}

For efficiently computing the gravitational potential on distributed-memory
systems, it is important to minimize the amout of data being explicitly
exchanged between compute tasks. The easiest parallel solution of the
aforementioned equations would be to collect the entire density field from
all computing units, calculate the gravitational potential in serial, and send the
result back to all taks.  This, however, would require a large amount of data
being communicated very inefficiently. More precisely, roughly $n_R \times
n_\vartheta \times n_\varphi \times 2$ floating-point numbers would be sent
serially,
where the factor of $2$ accounts for the collection of the data and their
subsequent re-distribution.  The serial calculation of the gravitational
potential would cause an additional load imbalance between the tasks.
Especially if the number of grid cells is very large, this strategy is
unrewarding.

Here, we present an approach for solving the gravitational potential accurately
on a spherical grid, whose angular domain is decomposed among a large number of
tasks, while the full radial slice is retained. This is precisely the setup used
in our neutrino-hydrodynamics code \textsc{Prometheus-Vertex}. If we rewrite
Eqs.\ (\ref{eq:phiin_tot}) and (\ref{eq:phiout_tot}) as
\begin{align}
    \Phi_{\textrm{in}}^{(\ell m)}(\myvector{r}) &= \frac{1}{r^{\ell+1}}
    \sum_{i=1}^{n_r} \mathcal{R}_{\textrm{in},i}^{(\ell)}
    \left(\cos(m\varphi)\,A_{\mathrm{C},i}^{(\ell m)} +
    \sin(m\varphi)\,A_{\mathrm{S},i}^{(\ell m)}\right),
    \label{eq:phiin} \\
    \Phi_{\textrm{out}}^{(\ell m)}(\myvector{r}) &= r^{\ell}
    \sum_{i=n_r+1}^{n_R} \mathcal{R}_{\textrm{out},i}^{(\ell)}
    \left(\cos(m\varphi)\,A_{\mathrm{C},i}^{(\ell m)} +
    \sin(m\varphi)\,A_{\mathrm{S},i}^{(\ell m)}\right),
    \label{eq:phiout}
\end{align}
and define
\begin{align}
    A_{\mathrm{C},i}^{(\ell m)} &\coloneqq \sum_{j=1}^{n_{\vartheta}}\sum_{k=1}^{n_\varphi}
    \varrho_{ijk}\,\mathcal{T}_j^{(\ell m)}\,\mathcal{C}_k^{(m)},
    \label{eq:ac} \\
    A_{\mathrm{S},i}^{(\ell m)} &\coloneqq \sum_{j=1}^{n_{\vartheta}}\sum_{k=1}^{n_\varphi}
    \varrho_{ijk}\,\mathcal{T}_j^{(\ell m)}\,\mathcal{S}_k^{(m)},
    \label{eq:as}
\end{align}
we immediately see that the summation over the angular domain is separated from
the remaining calculation.  After each compute task has solved Eqs.\
(\ref{eq:ac}) and (\ref{eq:as}) locally, a global summation of
$A_{\mathrm{C},i}^{(\ell m)}$ and $A_{\mathrm{S},i}^{(\ell m)}$ is performed.
This is the only step, where explicit data communication is necessary.  Parallel
computing libraries provide efficient built-in methods to compute the sum across
$n_\mathrm{tasks}$ compute tasks. In case of the commonly used ``Message Passing
Interface'' (MPI), such a function is \texttt{MPI\_Allreduce}, which is usually
an implementation of the ``recursive doubling'' algorithm using $\log_2
n_\mathrm{tasks}$ communication steps to compute the global sum
\citep{Thakur2005}.

With our method, we thus communicate $2\times n_R \times
\frac{1}{2}\left[(\ell_\mathrm{max} + 1)^2 + (\ell_\mathrm{max} + 1)\right]
\times \log_2 n_\mathrm{tasks}$ floating-point numbers serially, where the
factor of $2$ accounts for the two sums, $A_{\mathrm{C},i}^{(\ell m)}$ and
$A_{\mathrm{S},i}^{(\ell m)}$, and the third factor counts the number of
different terms for $0 \leq \ell \leq \ell_\mathrm{max}$ and $0 \leq m \leq
\ell$. The fourth factor is based on the assumption that the ``recursive
doubling'' algorithm is used implicitly by the parallel computing library as
explained above.
Considering a computational mesh with an angular resolution of two degrees, our
procedure reduces the amount of data being sent as long as $\ell_\mathrm{max}
\lesssim 55$. In the core-collapse supernova context, a realistic setting of
$\ell_\mathrm{max}$ would be considerably smaller than this threshold (e.g.,
$\sim\!25$), because the gravitational potential is dominated by the
proto-neutron star, whose density distribution is spherically symmetric to good
approximation.

The global summation of $A_{\mathrm{C},i}^{(\ell m)}$ and
$A_{\mathrm{S},i}^{(\ell m)}$ can be dominated by round-off errors, even in
\emph{double} floating-point precision. The reproducibility of the results is
not guaranteed, because the summation order depends on the number of compute
tasks and on implementation details of the message-passing library used for data
communication. To tackle this issue, the ``double-double summation'' algorithm
described by \citet{He2001} is applied, which treats each number as a pair of
the value itself and its error being accumulated during floating-point
operations. With this method, we achieve reproducibility of the results up to
machine precision.

Generally, only the terms that explicitly depend on the density have to be
calculated in each step of a time-dependent simulation. All other terms remain
constant and therefore need to be computed only once as long as the grid
coordinates do not change with time.

\section{Algorithm and its parallelization for the Yin-Yang grid}
\label{sec:yinyang}

\begin{figure}
    \centering
    \includegraphics[width=0.5\linewidth]{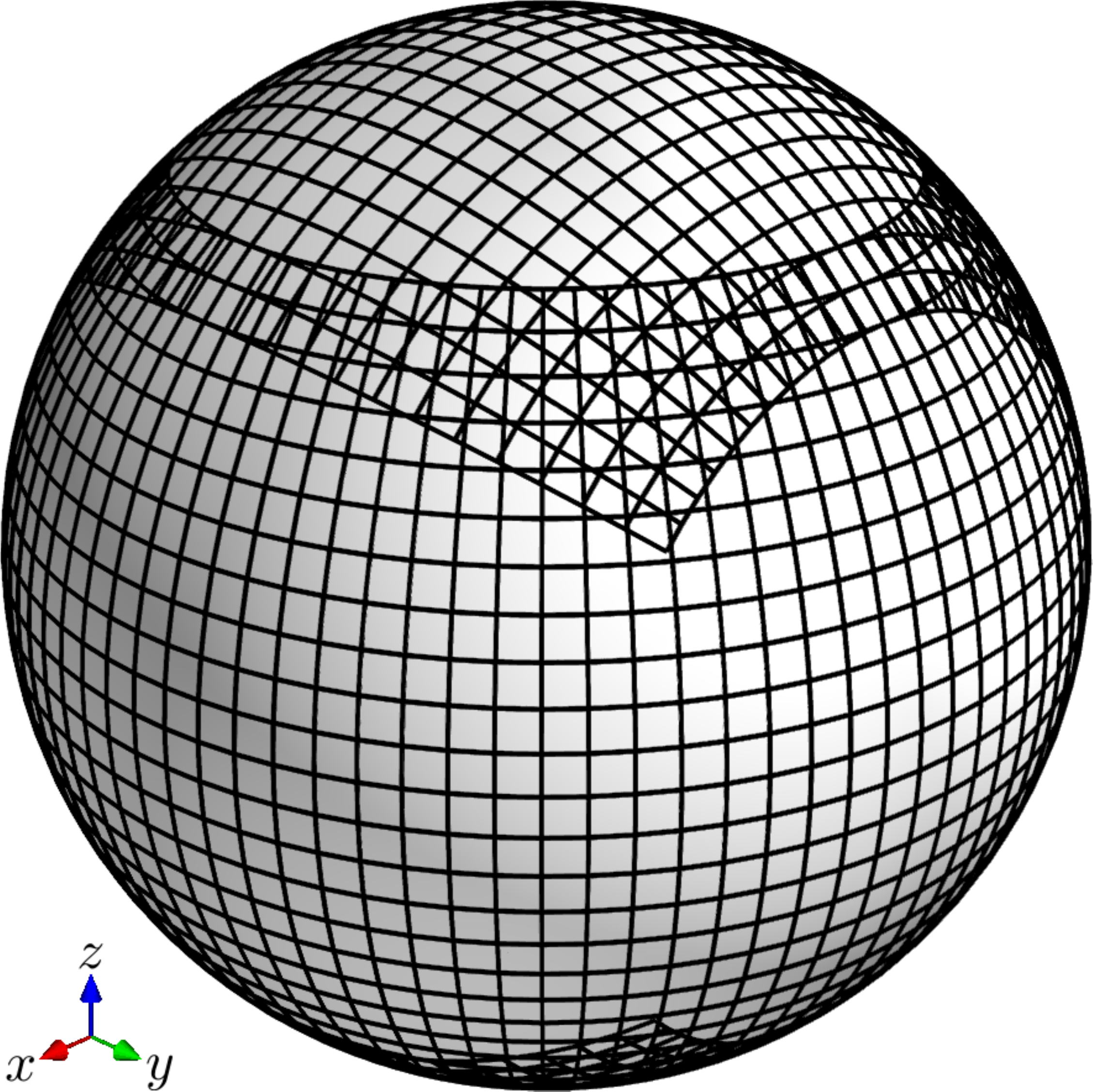}
    \caption{Three-dimensional visualization of the Yin-Yang grid cells on a
    sphere of constant radius.}
    \label{fig:yy}
\end{figure}

In its latest version, the \textsc{Prometheus-Vertex} code solves the
hydrodynamics on the Yin-Yang grid \citep{Kageyama2004,Wongwathanarat2010},
which is a combined mesh consisting of two low-latitude parts of a spherical
polar grid. Its main advantage over a spherical polar grid covering the full $4
\pi$ sphere is the exclusion of the grid singularities at the poles with their
numerical difficulties. We show the geometry of the Yin-Yang grid in Fig.\
\ref{fig:yy}.

Let the coordinates on \emph{Yin} and \emph{Yang} be denoted as ($r_\mathrm{Yin}$,
$\vartheta_\mathrm{Yin}$, $\varphi_\mathrm{Yin}$) and ($r_\mathrm{Yang}$,
$\vartheta_\mathrm{Yang}$, $\varphi_\mathrm{Yang}$), respectively, and defined
in the intervals
\begin{align}
    0 &\leq r \leq r_\mathrm{max},
    \nonumber \\
    \frac{\pi}{4}-\delta_\vartheta &\leq\vartheta\leq\frac{3\pi}{4}+\delta_\vartheta,
    \nonumber \\
    -\frac{3\pi}{4}-\delta_\varphi&\leq\varphi\leq\frac{3\pi}{4}+\delta_\varphi,
    \nonumber
\end{align}
where $r_\mathrm{max}$ is the outer radius of the grid. $\delta_\vartheta$ and
$\delta_\varphi$ are small values that control the width of the buffer zones in
the Yin-Yang overlap region. Both grid patches consist of $n_R \times
n_\vartheta \times n_\varphi$ cells and have equal coordinate axes.
The angular coordinates of a certain point on \emph{Yin} are transformed into
the \emph{Yang} system according to
\begin{align}
    \vartheta_\mathrm{Yang} &= \arccos\left(\sin\vartheta_\mathrm{Yin}
    \sin\varphi_\mathrm{Yin}\right),
    \label{eq:trafotheta} \\
    \varphi_\mathrm{Yang} &= \arctan\left(-\frac{\cos\vartheta_\mathrm{Yin}}
    {\sin\vartheta_\mathrm{Yin}\cos\varphi_\mathrm{Yin}}\right),
    \label{eq:trafophi}
\end{align}
where the inverse transformation is obtained by exchanging the subscripts.

The computation of the gravitational potential on the Yin-Yang grid is not
trivial, because both grid patches are rotated against each other and are
partially overlapping. \citet{Wongwathanarat2010} also used a version of
\textsc{Prometheus} with the Yin-Yang grid. They computed the gravitational
potential by interpolating the density from the individual Yin-Yang
grid parts onto a new ``auxiliary'' spherical polar grid that covers the entire
domain. The algorithm by \citet{Mueller1995} could then be applied without any
modifications. After that, the gravitational potential was mapped back to the
Yin-Yang grid patches.  Although the orientation and grid resolution of the
auxiliary grid is arbitrary, \citet{Wongwathanarat2010} aligned it with the
\emph{Yin} part for simplicity.

The method by \citet{Wongwathanarat2010}, however, has two major drawbacks for
us. Their code runs on shared-memory systems only, while our implementation of
the gravity solver should be optimized for distributed-memory architectures.
Collecting the density data from all compute tasks to a single worker would
be very inefficient due to the amount of explicit data communication and the
serialization of the calculation (see Section \ref{sec:parallel}). Furthermore,
interpolation errors of the density and the calculated gravitational
potential occur on the grid patch that is not aligned with the auxiliary spherical polar
grid (\emph{Yang}), while the data on the \emph{Yin} part can simply be copied.
This causes an artificial asymmetry between the Yin-Yang grid patches.

In order to efficiently compute the gravitational potential on the parallelized
Yin-Yang grid, we developed a new method based on Eqs.\ (\ref{eq:ac}) and
(\ref{eq:as}). It benefits from the linearity of integrals, i.e., integrals over
the entire computational domain can be written as a sum of two integrals
computed separately on both \emph{Yin} and \emph{Yang}. Only the overlap regions
between the grid patches have to be treated differently by introducing a surface
weight $w_{jk}$ being $0.5$ for grid cells inside the overlap and $1$ for cells
outside. The weight factor for cells that are intersected by a boundary line and
therefore only partially overlapped is calculated numerically \citep{Peng2006}.
The sum of all surface elements multiplied with the surface
weight yields
\begin{equation}
    \sum_{j=1}^{n_\vartheta} \sum_{k=1}^{n_\varphi}
    (\cos\vartheta_j^--\cos\vartheta_j^+) (\varphi_k^+-\varphi_k^-) \, w_{jk} = 2 \pi.
\end{equation}
Note that this summation is done on one grid patch (either \textit{Yin} or
\textit{Yang}), which therefore yields $2 \pi$.

Similar to the parallelization procedure described in Section
\ref{sec:parallel} being valid for spherical polar grids, our new approach for
the Yin-Yang grid is based on computing $A_{\mathrm{C},i}^{(\ell m)}$ and
$A_{\mathrm{S},i}^{(\ell m)}$ locally, followed by a global summation over all
compute tasks. In the Yin-Yang setup, each task operates on data either from
\emph{Yin} or \emph{Yang}, i.e., it calculates the following sums locally,
\begin{align}
    A_{\mathrm{C},i,\mathrm{Yin/Yang}}^{(\ell m)} &= \sum_{j=1}^{n_{\vartheta}}\sum_{k=1}^{n_\varphi}
    \varrho_{ijk,\mathrm{Yin/Yang}}\,w_{jk}\, \mathcal{T}_j^{(\ell m)}\,\mathcal{C}_k^{(m)},\\
    A_{\mathrm{S}, i, \mathrm{Yin/Yang}}^{(\ell m)} &= \sum_{j=1}^{n_{\vartheta}}\sum_{k=1}^{n_\varphi}
    \varrho_{ijk,\mathrm{Yin/Yang}}\,w_{jk}\,\mathcal{T}_j^{(\ell m)}\,\mathcal{S}_k^{(m)}.
\end{align}
Note that due to the symmetry between both grid patches and their identical
\emph{local} coordinates, the values of $w_{jk}$, $\mathcal{T}_j^{(\ell m)}$,
$\mathcal{C}_k^{(m)}$, and $\mathcal{S}_k^{(m)}$ are equal on \emph{Yin} and
\emph{Yang}.

After evaluating the latter equations, the global sums are computed separately
for both grid parts resulting in different values for
$A_{\mathrm{C},i,\mathrm{Yin}}^{(\ell m)}$,
$A_{\mathrm{S},i,\mathrm{Yin}}^{(\ell m)}$,
$A_{\mathrm{C},i,\mathrm{Yang}}^{(\ell m)}$, and
$A_{\mathrm{S},i,\mathrm{Yang}}^{(\ell m)}$. Four different potentials are
calculated using Eqs.\ (\ref{eq:phiin}), (\ref{eq:phiout}), and
(\ref{eq:potential}), namely $\Phi_\mathrm{Yin}(\myvector{r}_\mathrm{Yin})$,
$\Phi_\mathrm{Yin}(\myvector{r}_\mathrm{Yang})$,
$\Phi_\mathrm{Yang}(\myvector{r}_\mathrm{Yin})$, and
$\Phi_\mathrm{Yang}(\myvector{r}_\mathrm{Yang})$. Here, the subscript of $\Phi$
denotes the grid part on which $A_{\mathrm{C},i}^{(\ell m)}$ and
$A_{\mathrm{S},i}^{(\ell m)}$ are computed, while the subscript of $\myvector{r}$
determines in which coordinate system the angles $\vartheta$ and $\varphi$ in
Eqs.\ (\ref{eq:potential}), (\ref{eq:phiin}), and (\ref{eq:phiout}) are
expressed.

The final gravitational potential for all compute tasks on \emph{Yin} and
\emph{Yang} is, respectively,
\begin{align}
    \Phi(\myvector{r}_\mathrm{Yin}) &= \Phi_\mathrm{Yin}(\myvector{r}_\mathrm{Yin}) +
    \Phi_\mathrm{Yang}(\myvector{r}_\mathrm{Yin}), \\
    \Phi(\myvector{r}_\mathrm{Yang}) &= \Phi_\mathrm{Yin}(\myvector{r}_\mathrm{Yang}) +
    \Phi_\mathrm{Yang}(\myvector{r}_\mathrm{Yang}).
\end{align}
Hence, the contributions from both grid parts are transformed into the local
coordinates of the task and added. This procedure is completely symmetric
with respect to both Yin-Yang grid patches and does not introduce a preferred
direction. An interpolation onto an auxiliary grid is not required thus avoiding
further numerical errors.

\begin{figure}
    \centering
    \includegraphics[width=0.95\linewidth]{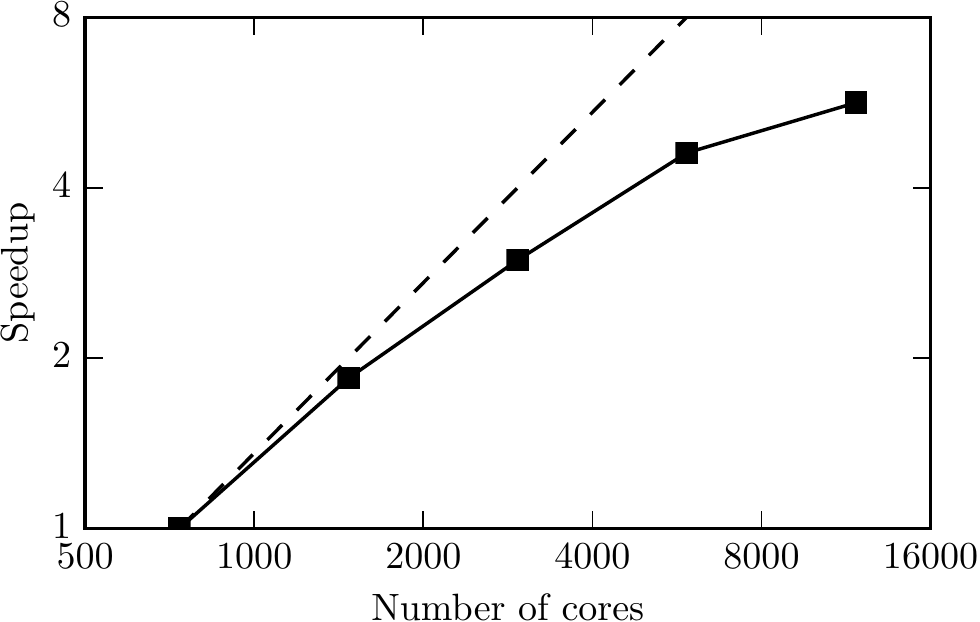}
    \caption{Strong scaling of the three-dimensional gravity solver on the
    Yin-Yang grid with an angular resolution of one degree. The dashed line
    indicates the theoretically perfect scaling behavior. Note the logarithmic
    scales of both coordinate axes.}
    \label{fig:scaling}
\end{figure}

In Fig.\ \ref{fig:scaling}, we present the strong scaling behavior of our
implementation of the Yin-Yang gravity solver. The gravitational potential was
computed for a density distribution on the Yin-Yang grid with an angular
resolution of one degree. While the grid configuration was kept constant, we
altered the number of processors and measured the speedup given as the current
wall-clock time normalized to the wall-clock time of the slowest run. The strong
scaling deviates from the theoretically expected perfect behavior with
increasing number of processors, because communication operations start to
dominate. However, we emphasize that the computation of the gravitational
potential by one single compute tasks as implemented in the
unparallelized code by \citet{Wongwathanarat2010} would result in a flat curve in Fig.\
\ref{fig:scaling}. Our approach is thus suitable for large-scale simulations
involving thousands of processors.

\section{Test calculations}
\label{sec:tests}

In order to validate our approach for solving the gravitational potential on the
Yin-Yang grid, we performed various tests. First, the gravitational potential of
a spherically symmetric density distribution was verified to be equal to the
corresponding analytical solution up to machine precision. Second, an
axisymmetric density distribution was mapped to the Yin-Yang grid with one degree
angular resolution and its gravitational potential was compared to the result
gained from an axisymmetric gravity solver\footnote{The axisymmetric solver has already been
used, e.g., by \citet{Buras2006a}, \citet{Marek2009}, and \citet{Summa2016}.}
yielding accurate agreement of $10^{-4}$. Third, the potential of a homogeneous
ellipsoid was compared to the semi-analytical solution. We will elaborate on the
latter test here in detail.

We constructed an ellipsoid of constant density $\varrho_0$ with semi-axes
$a$, $b$, and $c$ with $b = 1.5a$ and $c=2a$. In Cartesian coordinates, a point
is inside the ellipsoid if
\begin{equation}
    \left(\frac{x}{a}\right)^2 + \left(\frac{y}{b}\right)^2 +
    \left(\frac{z}{c}\right)^2 \leq 1.
\end{equation}
The gravitational potential of this configuration is given by
\citep{Chandrasekhar1969}
\begin{equation}
    \Phi(\myvector{r}) = \pi G \, \varrho_0 \, a b c \left(A(\myvector{r})
    \,x^2+B(\myvector{r})\,y^2+C(\myvector{r})\,z^2-D(\myvector{r})\right),\\
\end{equation}
where $\myvector{r}=(x,y,z)$ and
\begin{align}
    A(\myvector{r}) &\coloneqq \int_{u_0(\myvector{r})}^{\infty}\mathrm{d}u \left[(a^2+u)^3(b^2+u)(c^2+u)\right]^{-1/2},\\
    B(\myvector{r}) &\coloneqq \int_{u_0(\myvector{r})}^{\infty}\mathrm{d}u \left[(a^2+u)(b^2+u)^3(c^2+u)\right]^{-1/2},\\
    C(\myvector{r}) &\coloneqq \int_{u_0(\myvector{r})}^{\infty}\mathrm{d}u \left[(a^2+u)(b^2+u)(c^2+u)^3\right]^{-1/2},\\
    D(\myvector{r}) &\coloneqq \int_{u_0(\myvector{r})}^{\infty}\mathrm{d}u
    \left[(a^2+u)(b^2+u)(c^2+u)\right]^{-1/2}.
\end{align}
If the point $\myvector{r}$ is located inside the ellipsoid, $u_0(\myvector{r})
= 0$. Otherwise, $u_0(\myvector{r})$ is given as the positive root of the
equation
\begin{equation}
    \frac{x^2}{a^2+u_0} + \frac{y^2}{b^2+u_0} + \frac{z^2}{c^2+u_0} = 1.
\end{equation}
We evaluated the root and the integrals numerically up to a precision of
$10^{-13}$.

The transformation from the \emph{Yin} spherical polar coordinates to Cartesian
coordinates is given by
\begin{align}
    x &= r \sin\vartheta_\mathrm{Yin} \cos(\varphi_\mathrm{Yin}+\Delta\varphi), \\
    y &= r \sin\vartheta_\mathrm{Yin} \sin(\varphi_\mathrm{Yin}+\Delta\varphi), \\
    z &= r \cos\vartheta_\mathrm{Yin},
\end{align}
where we have introduced a rotation angle $\Delta\varphi=\pi/8$ in order to
break the symmetry between both Yin-Yang grid patches. To compute the Cartesian
coordinates for \emph{Yang}, the transformation given by Eqs.\
(\ref{eq:trafotheta}) and (\ref{eq:trafophi}) has to be applied to express the
\emph{Yang} coordinates in the \emph{Yin} system.

To estimate the correctness of our three-dimensional gravity solver for the case
of the homogeneous ellipsoid, we compared the calculated gravitational potential
in the entire computational domain (i.e., inside and outside the ellipsoid)
to the semi-analytical solution. This was done on the Yin-Yang grid, but also on
the standard spherical polar grid covering the whole sphere. Both mesh
configurations had an equidistant radial spacing and an outer radius
$r_\mathrm{max} = c$.  We varied both the grid resolution and the maximum
multipole order $\ell_\mathrm{max}$. In the following, we denote the models with
$n_R=400$ and an angular resolution of two degrees as ``low resolution'' (LR). The
runs with twice as many grid points in all directions are referred to as ``high
resolution'' (HR).

\begin{figure}
    \centering
    \includegraphics[width=0.95\linewidth]{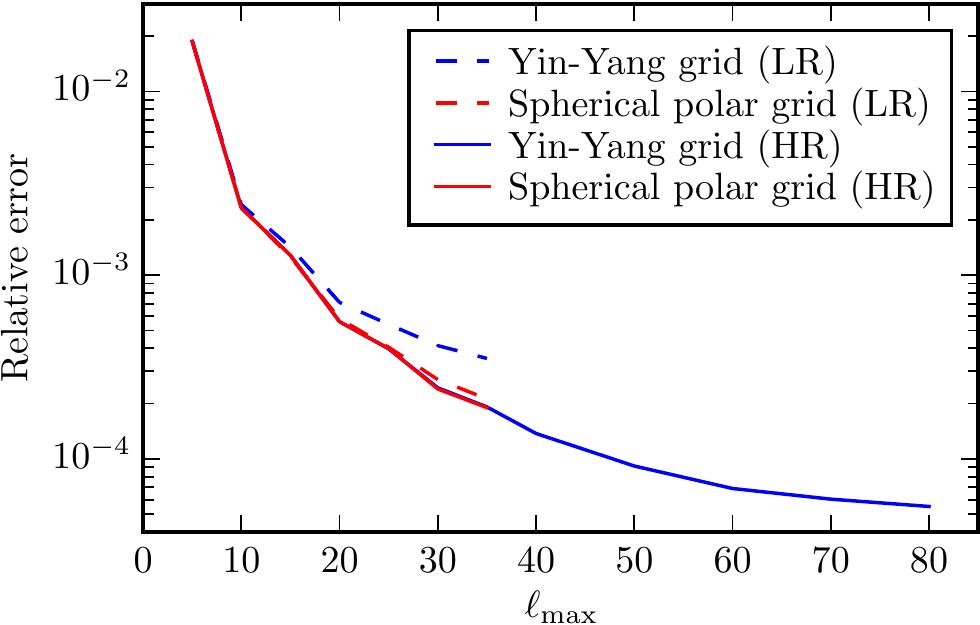}
    \caption{Maximum relative error of the calculated gravitational potential of
    the homogeneous ellipsoid compared to the semi-analytical solution as a function
    of maximum multipole order $\ell_\mathrm{max}$. The results computed on the
    spherical polar grid and on the Yin-Yang grid are shown with \emph{red} and
    \emph{blue} lines, respectively. \emph{Dashed} lines mark ``low-resolution''
    (LR) setups with $n_R=400$ and a two degree angular resolution, while
    \emph{solid} lines show ``high-resolution'' (HR) grids with $n_R=800$ and a
    one degree angular resolution.}
    \label{fig:ellipsoid1}
\end{figure}

In Fig.\ \ref{fig:ellipsoid1}, we show the maximum relative error of the
calculated gravitational potential with respect to the semi-analytical solution for
the different mesh configurations and grid resolutions. As expected, the
relative errors of the HR models are always smaller than in the LR cases, because
the gravitational potential is sampled more accurately in the former runs and
the surface of the ellipsoid is better resolved. While for the LR case the results
computed on the Yin-Yang grid and on the spherical polar grid deviate from each
other, both setups yield basically identical solutions in the HR case. This demonstrates that
our adaption of the gravity solver performs correctly on the Yin-Yang grid.

With increasing maximum multipole order $\ell_\mathrm{max}$, the relative error
decreases in all setups, since the shape of the ellipsoid is approximated more
accurately. Ideally, an infinite number of multipole terms would be required to
precisely reproduce an ellipsoidal configuration. Including more higher-order
terms thus reduces the deviation from the semi-analytical solution.

\begin{figure}
    \centering
    \includegraphics[width=0.95\linewidth]{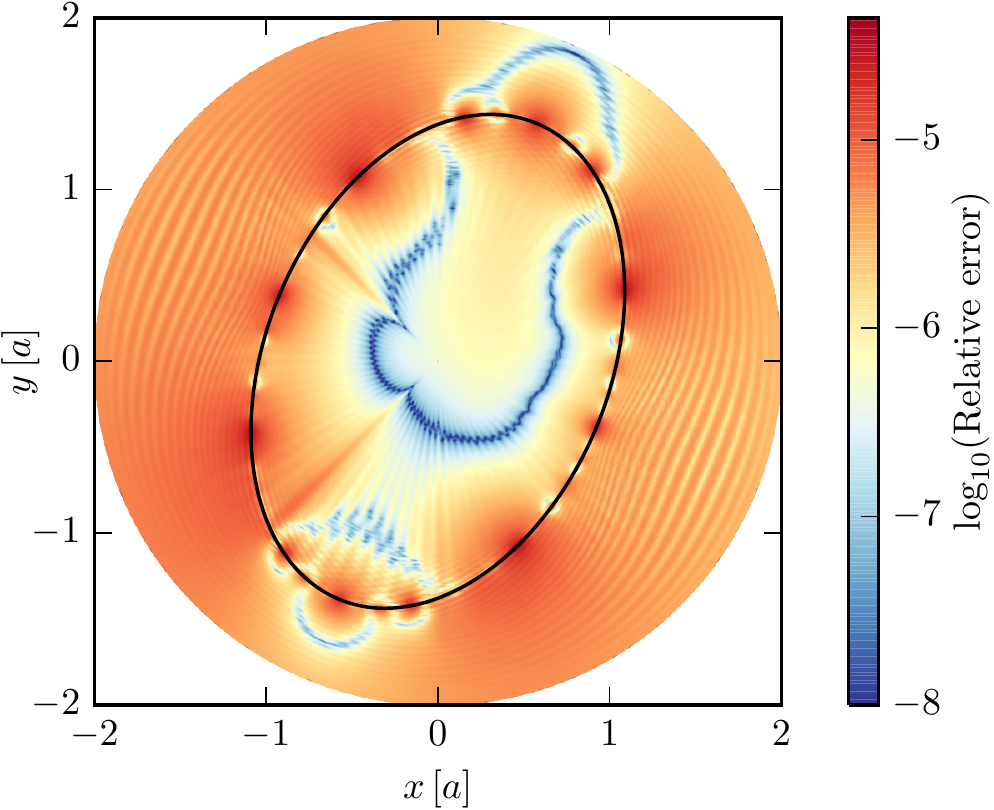}
    \caption{Relative error of the calculated gravitational potential in a
    cross-sectional plane through the homogeneous ellipsoid on the Yin-Yang grid
    compared to the semi-analytical
    solution. The data are displayed with a logarithmic color coding in the
    $x$-$y$ plane ($z=0$) for the ``high-resolution'' (HR) case with
    $\ell_\mathrm{max}=80$. The surface of the ellipsoid is indicated with a
    black ellipse. Note that the entire computational domain is shown.}
    \label{fig:ellipsoid2}
\end{figure}

The spatial distribution of the relative error on the Yin-Yang grid is shown in
Fig.\ \ref{fig:ellipsoid2} for the HR case with $\ell_\mathrm{max}=80$ in the
plane $z=0$. It can
clearly be seen that the error is dominated by the cells close to the surface of
the ellipsoid. Their deviation from the semi-analytical solution is two to three
orders of magnitude larger than for cells near the grid origin. Machine
precision is not reached in the center, because the gravitational potential
there is also dependent on the density distribution outside (see Eq.\ (\ref{eq:potential}) and the
term $\Phi_\mathrm{out}^{(\ell m)}$ there). Since the ellipsoidal surface is not sampled to
machine precision, we do not expect this either for the gravitational potential
in the central region.

\section{Discussion and Conclusions}
\label{sec:discussion}

In this work, we have presented a new method for solving the three-dimensional
gravitational potential of a density field on the Yin-Yang grid based on the
algorithm developed by \citet{Mueller1995}. Our approach is particularly
efficient on distributed-memory machines with a large number of compute tasks,
because the amount of data being explicitly communicated is minimized and the
work load is evenly distributed across all tasks resulting in a good scaling
behavior.

In contrast to the strategy by \citet{Wongwathanarat2010}, who interpolated the
density from the Yin-Yang grid onto an auxiliary spherical mesh, we perform all
operations on the original Yin-Yang grid. Our algorithm is completely symmetric
with respect to the Yin-Yang grid patches and does not introduce a preferred
direction. The validity of the results have been verified by performing test
calculations that have an analytical or a semi-analytical solution.

Our ellipsoidal test configuration mimics the situation near the surface of the
newly formed neutron star during the later post-bounce evolution of a
core-collapse supernova.  The increasingly steeper density gradient between the
neutron star, which may be centrifugally deformed due to rapid rotation, and its
surrounding medium is represented by the boundary of the ellipsoid. Since our
solver can accurately determine the gravitational potential in this considered
test setup with its extremely ``sharp'' surface, proper results can also be
expected for the more moderate situation met during the birth of neutron stars
during stellar core collapse and explosion.

The maximum multipole order $\ell_\mathrm{max}$ of the spherical harmonics
decomposition determines the accuracy of the gravitational potential. The
computational cost increases roughly as $\ell_\mathrm{max}^2$. A suitable
balance between accuracy and computing time has to be found depending on the
characteristics of the investigated physical problem. In core-collapse supernova
models, the gravitational potential is dominated by the approximately
spherically symmetric proto-neutron star. Therefore a moderate
$\ell_\mathrm{max}$ of $\sim\!25$ should be able to capture most of the
asymmetries.

\acknowledgments
\section*{Acknowledgements}
We thank Alexander Summa and Annop Wongwathanarat for fruitful discussions.
This project was supported by the European Research Council through grant
ERC-AdG No.\ 341157-COCO2CASA and by the Deutsche Forschungsgemeinschaft through
the Excellence Cluster ``Universe'' (EXC-153).  The test calculations were
performed on the \emph{Hydra} system of the Max Planck Computing and Data
Facility (MPCDF). The code scaling was tested on \emph{SuperMUC} at the Leibniz
Supercomputing Center with resources granted by the Gauss Centre for
Supercomputing (LRZ project ID: pr53yi).

\software{\textsc{Prometheus-Vertex} \citep{Fryxell1989,Rampp2002,Buras2006a},
NumPy and Scipy \citep{Oliphant2007}, IPython \citep{Perez2007}, Matplotlib \citep{Hunter2007}.}

\appendix
\section{Recurrence relations for the integrals of the associated Legendre polynomials}
\label{appendix}

The integrals $\mathcal{T}_j^{(\ell m)}$ defined in Eq.\ (\ref{eq:tlm}) can be
calculated using the following recurrence relations \citep{Zwerger1995}, which
are evaluated in the order given below.
\begin{align}
    \mathcal{T}_j^{(0 0)} &= P_1^0(\cos\vartheta_j^{-}) - P_1^0(\cos\vartheta_j^{+}), \\
    \mathcal{T}_j^{(\ell 0)} &= \frac{1}{2\ell+1}\left[P_{\ell-1}^0(\cos\vartheta_j^{+}) -
    P_{\ell-1}^0(\cos\vartheta_j^{-}) + P_{\ell+1}^0(\cos\vartheta_j^{-}) -
    P_{\ell+1}^0(\cos\vartheta_j^{+})\right] \quad(\text{only for }\ell > 0), \\
    \mathcal{T}_j^{(11)} &= \frac{1}{2}\left[\vartheta_j^{-}-\vartheta_j^{+}+\frac{1}{2}
    \left(\sin(2\vartheta_j^{+}) - \sin(2\vartheta_j^{-})\right)\right], \\
    \mathcal{T}_j^{(22)} &= \frac{1}{4}\left(\cos(3\vartheta_j^{+})-\cos(3\vartheta_j^{-})\right)
    - 9\left(\cos\vartheta_j^{+} - \cos\vartheta_j^{-}\right), \\
    \mathcal{T}_j^{(\ell \ell)} &= \frac{1}{\ell+1}\left[\cos\vartheta_j^{-}\,
    P_\ell^\ell(\cos\vartheta_j^{-}) - \cos\vartheta_j^{+}\, P_\ell^\ell(\cos\vartheta_j^{+}) +
    \ell(2\ell-3)(2\ell-1)\,\mathcal{T}_j^{(\ell-2,\ell-2)}\right], \\
    \mathcal{T}_j^{(\ell,\ell-1)} &= -\frac{1}{\ell+1}\left[\sin\vartheta_j^{+}\,
    P_\ell^\ell(\cos\vartheta_j^{+}) - \sin\vartheta_j^{-}\,P_\ell^\ell(\cos\vartheta_j^{-})\right], \\
    \mathcal{T}_j^{(\ell m)} &= \frac{1}{(\ell+1)(\ell-m)}\left[(2\ell-1)
    \left(\sin^2\vartheta_j^{+} \,P_{\ell-1}^m (\cos\vartheta_j^{+}) -
    \sin^2\vartheta_j^{-} \,P_{\ell-1}^m (\cos\vartheta_j^{-})\right) +
    (\ell-2)(\ell+m-1)\,\mathcal{T}_j^{(\ell-2,m)}\right].
\end{align}

\bibliographystyle{apj}
\bibliography{paper}

\begin{thebibliography}{}
\expandafter\ifx\csname natexlab\endcsname\relax\def\natexlab#1{#1}\fi

\bibitem[{{Buras} {et~al.}(2006){Buras}, {Rampp}, {Janka}, \&
  {Kifonidis}}]{Buras2006a}
{Buras}, R., {Rampp}, M., {Janka}, H.-T., \& {Kifonidis}, K. 2006, A\&A, 447,
  1049

\bibitem[{{Chandrasekhar}(1969)}]{Chandrasekhar1969}
{Chandrasekhar}, S. 1969, {Ellipsoidal figures of equilibrium} (Yale Univ.
  Press)

\bibitem[{{Falle} {et~al.}(2017){Falle}, {Vaidya}, \& {Hartquist}}]{Falle2017}
{Falle}, S.~A.~E.~G., {Vaidya}, B., \& {Hartquist}, T.~W. 2017, MNRAS, 465, 260

\bibitem[{{Fryxell} {et~al.}(1989){Fryxell}, {M{\"u}ller}, \&
  {Arnett}}]{Fryxell1989}
{Fryxell}, B., {M{\"u}ller}, E., \& {Arnett}, D. 1989, in Proceedings of the
  5$^\mathrm{th}$ Workshop on Nuclear Astrophysics, ed. W.~{Hillebrandt} \&
  E.~{M{\"u}ller}, 100

\bibitem[{{He} \& {Ding}(2001)}]{He2001}
{He}, Y., \& {Ding}, C. H.~Q. 2001, J. Supercomput., 18, 259

\bibitem[{{Hunter}(2007)}]{Hunter2007}
{Hunter}, J.~D. 2007, CSE, 9, 90

\bibitem[{{Kageyama} \& {Sato}(2004)}]{Kageyama2004}
{Kageyama}, A., \& {Sato}, T. 2004, GGG, 5, 9005

\bibitem[{{Komatitsch} \& {Tromp}(2002)}]{Komatitsch2002}
{Komatitsch}, D., \& {Tromp}, J. 2002, GeoJI, 150, 303

\bibitem[{{Lentz} {et~al.}(2015){Lentz}, {Bruenn}, {Hix}, {Mezzacappa},
  {Messer}, {Endeve}, {Blondin}, {Harris}, {Marronetti}, \&
  {Yakunin}}]{Lentz2015}
{Lentz}, E.~J., {Bruenn}, S.~W., {Hix}, W.~R., {et~al.} 2015, ApJL, 807, L31

\bibitem[{{Liebend{\"o}rfer} {et~al.}(2005){Liebend{\"o}rfer}, {Rampp},
  {Janka}, \& {Mezzacappa}}]{Liebendoerfer2005a}
{Liebend{\"o}rfer}, M., {Rampp}, M., {Janka}, H.-T., \& {Mezzacappa}, A. 2005,
  \apj, 620, 840

\bibitem[{{Marek} {et~al.}(2006){Marek}, {Dimmelmeier}, {Janka}, {M{\"u}ller},
  \& {Buras}}]{Marek2006}
{Marek}, A., {Dimmelmeier}, H., {Janka}, H.-T., {M{\"u}ller}, E., \& {Buras},
  R. 2006, A\&A, 445, 273

\bibitem[{{Marek} \& {Janka}(2009)}]{Marek2009}
{Marek}, A., \& {Janka}, H.-T. 2009, ApJ, 694, 664

\bibitem[{{Melson} {et~al.}(2015{\natexlab{a}}){Melson}, {Janka}, {Bollig},
  {Hanke}, {Marek}, \& {M{\"u}ller}}]{Melson2015b}
{Melson}, T., {Janka}, H.-T., {Bollig}, R., {et~al.} 2015{\natexlab{a}}, ApJL,
  808, L42

\bibitem[{{Melson} {et~al.}(2015{\natexlab{b}}){Melson}, {Janka}, \&
  {Marek}}]{Melson2015a}
{Melson}, T., {Janka}, H.-T., \& {Marek}, A. 2015{\natexlab{b}}, ApJL, 801, L24

\bibitem[{{Mondal} \& {Korenaga}(2018)}]{Mondal2018}
{Mondal}, P., \& {Korenaga}, J. 2018, GeoJI, 212, 1859

\bibitem[{{M{\"u}ller} \& {Chan}(2018)}]{Mueller2018}
{M{\"u}ller}, B., \& {Chan}, C. 2018, ArXiv e-prints, 1806.06623

\bibitem[{{M{\"u}ller} {et~al.}(2010){M{\"u}ller}, {Janka}, \&
  {Dimmelmeier}}]{Mueller2010}
{M{\"u}ller}, B., {Janka}, H.-T., \& {Dimmelmeier}, H. 2010, \apjs, 189, 104

\bibitem[{{M{\"u}ller} {et~al.}(2017){M{\"u}ller}, {Melson}, {Heger}, \&
  {Janka}}]{Mueller2017}
{M{\"u}ller}, B., {Melson}, T., {Heger}, A., \& {Janka}, H.-T. 2017, MNRAS,
  472, 491

\bibitem[{{M{\"u}ller} \& {Steinmetz}(1995)}]{Mueller1995}
{M{\"u}ller}, E., \& {Steinmetz}, M. 1995, CoPhC, 89, 45

\bibitem[{{Oliphant}(2007)}]{Oliphant2007}
{Oliphant}, T.~E. 2007, CSE, 9, 10

\bibitem[{{Ott} {et~al.}(2018){Ott}, {Roberts}, {da Silva Schneider}, {Fedrow},
  {Haas}, \& {Schnetter}}]{Ott2018}
{Ott}, C.~D., {Roberts}, L.~F., {da Silva Schneider}, A., {et~al.} 2018, \apjl,
  855, L3

\bibitem[{{Peng} {et~al.}(2006){Peng}, {Xiao}, \& {Takahashi}}]{Peng2006}
{Peng}, X., {Xiao}, F., \& {Takahashi}, K. 2006, QJRMS, 132, 979

\bibitem[{{Perez} \& {Granger}(2007)}]{Perez2007}
{Perez}, F., \& {Granger}, B.~E. 2007, CSE, 9, 21

\bibitem[{{Rampp} \& {Janka}(2002)}]{Rampp2002}
{Rampp}, M., \& {Janka}, H.-T. 2002, A\&A, 396, 361

\bibitem[{{Roberts} {et~al.}(2016){Roberts}, {Ott}, {Haas}, {O'Connor},
  {Diener}, \& {Schnetter}}]{Roberts2016}
{Roberts}, L.~F., {Ott}, C.~D., {Haas}, R., {et~al.} 2016, \apj, 831, 98

\bibitem[{{Simon} {et~al.}(2016){Simon}, {Armitage}, {Li}, \&
  {Youdin}}]{Simon2016}
{Simon}, J.~B., {Armitage}, P.~J., {Li}, R., \& {Youdin}, A.~N. 2016, ApJ, 822,
  55

\bibitem[{{Summa} {et~al.}(2016){Summa}, {Hanke}, {Janka}, {Melson}, {Marek},
  \& {M{\"u}ller}}]{Summa2016}
{Summa}, A., {Hanke}, F., {Janka}, H.-T., {et~al.} 2016, ApJ, 825, 6

\bibitem[{{Summa} {et~al.}(2018){Summa}, {Janka}, {Melson}, \&
  {Marek}}]{Summa2018}
{Summa}, A., {Janka}, H.-T., {Melson}, T., \& {Marek}, A. 2018, ApJ, 852, 28

\bibitem[{{Takiwaki} {et~al.}(2012){Takiwaki}, {Kotake}, \&
  {Suwa}}]{Takiwaki2012}
{Takiwaki}, T., {Kotake}, K., \& {Suwa}, Y. 2012, ApJ, 749, 98

\bibitem[{{Takiwaki} {et~al.}(2014){Takiwaki}, {Kotake}, \&
  {Suwa}}]{Takiwaki2014}
---. 2014, ApJ, 786, 83

\bibitem[{{Thakur} {et~al.}(2005){Thakur}, {Rabenseifner}, \&
  {Gropp}}]{Thakur2005}
{Thakur}, R., {Rabenseifner}, R., \& {Gropp}, W. 2005, IJHPCA, 19, 49

\bibitem[{{Wongwathanarat} {et~al.}(2010){Wongwathanarat}, {Hammer}, \&
  {M{\"u}ller}}]{Wongwathanarat2010}
{Wongwathanarat}, A., {Hammer}, N.~J., \& {M{\"u}ller}, E. 2010, A\&A, 514, A48

\bibitem[{{Zwerger}(1995)}]{Zwerger1995}
{Zwerger}, T. 1995, PhD thesis, Technische Universit{\"a}t M{\"u}nchen

\end{thebibliography}

\end{document}